\documentclass[aps,prb,twocolumn,superscriptaddress]{revtex4-1}

\usepackage{graphicx}
\usepackage{color}
\usepackage{here}

\begin{document}

\title{Enhanced thermoelectricity by controlled local structure in bismuth-chalcogenides}
% Force line breaks with \\

\author{K. Terashima}
\email[]{Author to whom correspondence should be addressed.  Electronic mail: k-terashima@cc.okayama-u.ac.jp}
%\thanks{}
\affiliation{Research Institute for Interdisciplinary Science, Okayama University, Okayama, 700-8530, Japan}

\author{Y. Yano}
\affiliation{Graduate School of Natural Sciences, Okayama University, Okayama, 700-8530, Japan}
\author{E. Paris}
\altaffiliation{Present address: Research Department Synchrotron Radiation and Nanotechnology, Paul Scherrer Institut, CH-5232 Villigen PSI, Switzerland}
\affiliation{Dipartimento di Fisica, Universit\'{a} di Roma ``La Sapienza'' - P. le Aldo Moro 2, 00185, Roma, Italy }
\author{Y. Goto}
\affiliation{Department of Physics, Tokyo Metropolitan University, Tokyo 192-0397, Japan}
\author{Y. Mizuguchi}
\affiliation{Department of Physics, Tokyo Metropolitan University, Tokyo 192-0397, Japan}
\author{Y. Kamihara}
\affiliation{Department of Applied Physics and Physico-Informatics, Keio University, Yokohama 223-8522, Japan}
\author{T. Wakita}
\affiliation{Research Institute for Interdisciplinary Science, Okayama University, Okayama, 700-8530, Japan}
\author{Y. Muraoka}
\affiliation{Research Institute for Interdisciplinary Science, Okayama University, Okayama, 700-8530, Japan}
\author{N. L. Saini}
\affiliation{Dipartimento di Fisica, Universit\'{a} di Roma ``La Sapienza'' - P. le Aldo Moro 2, 00185, Roma, Italy }
\author{T. Yokoya}
\affiliation{Research Institute for Interdisciplinary Science, Okayama University, Okayama, 700-8530, Japan}


\begin{abstract}
Spectroscopic techniques, including photoelectron spectroscopy, diffuse reflectance, and x-ray absorption, are used to investigate the electronic structure and the local structure of LaOBiS$_{2-x}$Se$_x$ thermoelectric material.  It is found that Se substitution effectively suppresses local distortion, that can be responsible for the increased carrier mobility together with a change in the electronic structure.  The results suggest a possible way to control thermoelectric properties by tuning of the local crystal structure of these materials.
\end{abstract}

\maketitle

\section{INTRODUCTION}
	Thermoelectric materials with layered structures hold possibillities to modify their physical properties, $e.g.$, by partial substitutions in the block layer and/or conducting layer.  Among them, REOBiCh$_2$ (RE = rare earth and Ch = chalcogen) system \cite{Mizuguchi1st} is acquiring much attention due to its structural similarity with high-$T_c$ cuprates and iron-based superconductors in which carrier number and other physical parameters are known to be controlled by combinations of constituting block layers and conducting layers \cite{Ce, Pr, Nd, Eu, Sm, TM, Sr, HEA, inter, Se}. The parent material, LaOBiCh$_2$ is insulating, and an object of superconductivity research since it becomes superconductor by carrier doping \cite{Mizuguchi1st}. It was found \cite{JAP1} that isovalent substitution of chalcogen atoms in parent compound induces a large improvement of thermoelectric figure of merit given by $ZT$ = ($S^2$/$\rho \kappa)T$.  The $ZT$ in the LaOBiS$_{2-x}$Se$_x$ system has been found \cite{NishidaJPSJ} to be as high for $\sim$0.36 in {\it x} = 1.0 at 650 K.
	
The thermoelectric $ZT$ consists of several physical parameters, namely Seebeck coefficient ($S$), resistivity ($\rho$), and thermal conductivity ($\kappa$).  For $x$ = 1.0, the absolute value of Seebeck coefficient is $\sim$20 \% higher while the total  thermal conductivity remains similar to that of  $x$ = 0.0, instead the resistivity shows a significant decrease \cite{JAP1}.  It can be seen in Fig. 1 that the electrical resistivity of LaOBiSSe (corresponds to $x$ = 1.0) below room temperature is smaller than that of LaOBiS$_2$ by about a factor of 10. Judging from the reported relationship between the carrier concentration and Seebeck coefficient in LaOBiCh$_2$  \cite{Mizuguchicogent}, the carrier concentrations in LaOBiS$_2$ and LaOBiSSe are expected to be similar\cite{NishidaJPSJ}. In a previous study \cite{NishidaJPSJ}, it has been argued that such a difference in electrical resistivity could be related with large enhancement of carrier mobility by Se substitution, attributed to the chemical pressure effect or more precisely, the misfit strain effect in the layered structure since the ionic radius of Se is larger than that of S.  This may cause an enhanced overlap between adjacent orbitals through substitution, resulting in a change in the electronic structure. However, there is no direct experimental evidence providing microscopic mechanism responsible for improving the thermoelectric properties in this system. This is also important due to the fact that REOBiS$_2$ is structurally instable \cite{Sugimoto} with polymorphism \cite{poly1, poly2} characterized by coexisting atomic configurations and hence apart from the knowledge of the electronic transport, study of the local structure should provide important feedthrough.

\begin{figure}[h]
\includegraphics[width=8.3 cm]{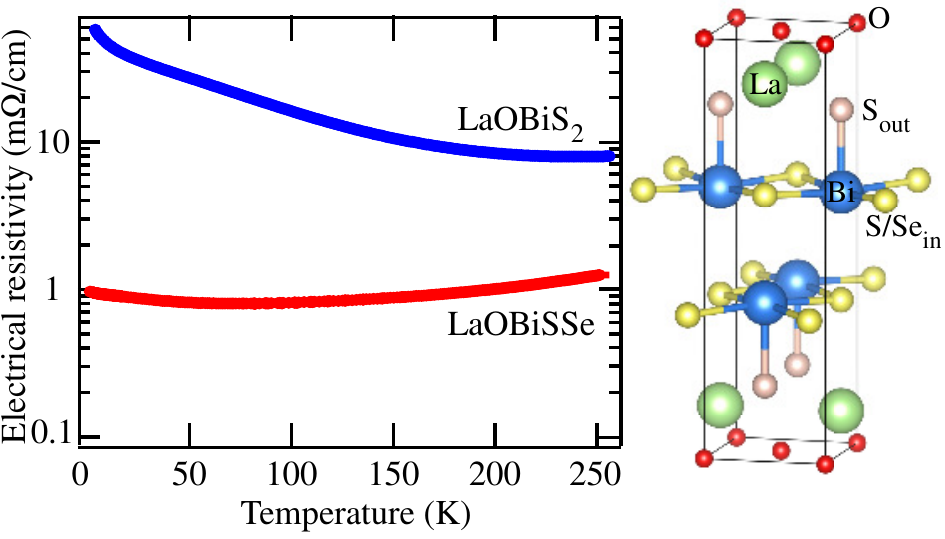}%
\caption{Left: Temperature dependence of electrical resistivity of LaOBiS$_2$ (blue) and LaOBiSSe (red). Right: Crystal structure of LaOBiCh$_2$.}
\end{figure}	

	Here, we have performed a comparative study of the electronic structure and local structure of LaOBiS$_2$ and LaOBiSSe using photoemission, diffuse reflectance, band calculation, and x-ray absorption fine structure, to address possible origin of the enhanced carrier mobility by Se substitution.  We have observed that valence band structure is altered by Se substitution with the band gap decreasing by about half, which agrees qualitatively with first principle band structure calculations.  According to band structure calculations, the top of valence band mainly consists of in-plane chalcogen {\it p}-orbitals which is strongly affected by partial substitution of chalcogen atoms.  Calculated conduction band does not show any substantial change with Se substitution, indicating that the large enhancement of carrier mobility below room temperature may not be accounted for the change in the electronic structure.  On the other hand, x-ray absorption fine structure shows marked change by substitution, indicating that the local distortion of in-plane Bi-Chalcogen plane in LaOBiSSe compound is drastically suppressed with respect to LaOBiS$_2$.  It has been suggested that this suppression of the local distortion might be playing a large contribution in enhanced thermoelectric property of this system.  
%We discuss the mechanism for controlling the distortion in light with the energy diagram of Bismuth and Chalcogen orbitals including Bi lone pair.

\section{METHODS}
	Polycrystalline LaOBiS$_2$ and LaOBiSSe samples were prepared by solid-state-reaction method\cite{Mizuguchi1st, JAP1}. Same procedure was used for preparing the two samples with the same person as the former reports of crystal structure and thermoelectric properties\cite{NishidaJPSJ, JAP2}.  Temperature dependence of electrical resistivity was measured by the four-probe methods.
% where Au wires were contacted to the sample with Ag paste.
	
	Photoelectron spectroscopy measurements on pelletized samples were performed using Scienta-Omicron R4000 analyzer constructed in Okayama University.   Xe I$\alpha$ ({\it h}$\nu$ = 8.437 eV), He I$\alpha$ (21.218 eV), and He II$\alpha$ (40.814 eV) lines were used to excite photoelectrons.  The energy resolution was set to 10 meV for Xe I and He I and 50 meV for He II, and the spectra were taken by the transmission mode in an ultrahigh vacuum of ~8$\times$10$^{-9}$ Pa.  The Fermi energy ($E_F$) of samples were calibrated by that of gold, electronically contacted with samples.  During the measurements, we found no signature of charging.  Clean surfaces for the measurements were obtained by {\it in situ} scraping of samples at ~2$\times$10$^{-8}$ Pa.

	Total diffuse reflectance spectra (R) were measured by a spectrometer with an integrating sphere (Hitachi High-Tech, U-4100). Al$_2$O$_3$ powders were used as the standard reference. Optical absorption coefficient ($\alpha$) was determined from the $R$ value via the Kubelka–Munk equation, $(1-R)^2/2R$ = $\alpha$/$s$, where $s$ is the scattering factor.
	
Bi L$_{3}$-edge ($E$=13418 eV) x-ray absorption measurements were used to probe the local structure of LaOBiS$_2$ and LaOBiSSe. The measurements were performed at the beamline BM26A of the European Synchrotron Radiation Facility (ESRF) where the synchrotron light was monochromatized using a double crystal Si(111) monochromator. The measurements were made sequentially at low temperature ($T$ = 30 K) in transmission mode. The finely powdered samples of LaOBiS$_2$ and LaOBiSSe were diluted uniformly in boron nitride matrix and pressed into pellets of 13 mm diameter, for obtaining the edge jump to be about 1. 
%Four different x-ray absorption scans were measured on each samples to ensure the reproducibility of the spectra and to obtain high signal to noise ratio. 
The EXAFS oscillations were extracted by the standard procedure based on the cubic spline fit to the pre-edge subtracted absorption spectrum \cite{Bunker}.

	First principles calculations were performed using WIEN2k code \cite{WIEN2k} where spin-orbit coupling was included.  The lattice parameters were taken from x-ray diffraction measurements on LaOBiS$_2$ and LaOBiSSe reported earlier \cite{JAP2}.  The modified Becke-Johnson (mBJ) potential proposed by Tran and Blaha \cite{mBJ1, mBJ2} was used as exchange correlation functional, as it has been discussed to be appropriate for BiS$_2$ system \cite{Ochi}.  For simplicity, we assumed the tetragonal crystal structure \cite{mono} despite a symmetry reduction in both compounds \cite{Athauda, Sagayama, Nagasaka}.  24$\times$24$\times$7 $k$-mesh was used with the $RK_{max}$ parameter being 7.

\begin{figure}
\includegraphics[width=8cm]{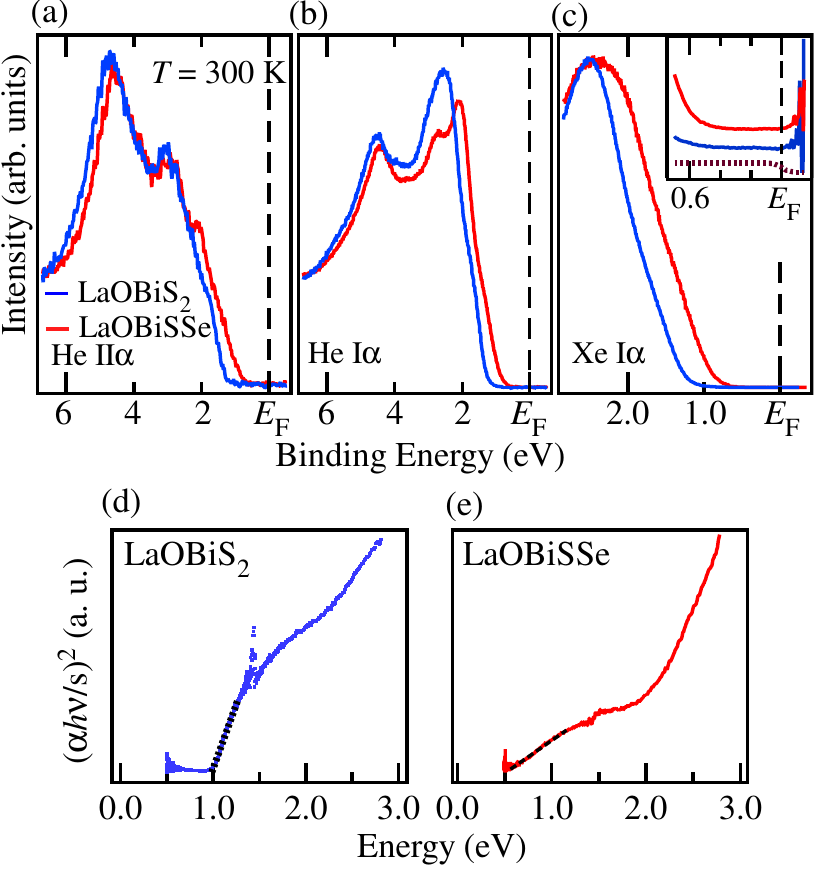}%
\caption{Photoemission spectra of LaOBiS$_2$ (blue) and LaOBiSSe (red) taken by (a) He II$\alpha$, (b) He I$\alpha$, and (c) Xe I$\alpha$ lines at $T$ = 300 K.  Inset in (c) shows the spectra near {\it E}$_{\rm F}$ divided by Fermi-Dirac function convoluted by experimental energy resolution (shown as dotted line).  (d), (e) ($\alpha${\it h}$\nu$/s)$^2$ as a function of {\it h}$\nu$ on LaOBiS$_2$ and LaOBiSSe, respectively. Dotted lines in (d) and (e) are guide for the eye.}
\end{figure}

\section{RESULTS AND DISCUSSION}												
	Firstly, we focus on the effect of Se substitution on the electronic structure. Figure 2 shows photoemission spectra of LaOBiS$_2$ and LaOBiSSe measured using several ultraviolet photon energies at $T$ = 300 K.  Spectra in Figs. 2(a) and 2(b) are normalized by total intensity over the shown area, while those in 2(c) are normalized by the peak height.  The overall width of the valence band and the energy position of the top of the valence band in LaOBiS$_2$ is consistent with our earlier soft x-ray photoemission work \cite{Nagira}.  Through S substitution with Se, we have observed that the top of the valence band becomes closer to $E_F$ at all the photon energies.  Inset of Fig. 2(c) shows the near-{\it E}$_{\rm F}$ spectra divided by Fermi-Dirac function of $T$ = 300 K convoluted by experimental energy resolution.  We have found that the {\it E}$_{\rm F}$s of both sample are located at the bottom of the conduction band, that would be consistent with the sign of the Seebeck coefficient.  Therefore the magnitude of the band gap can be evaluated for both samples, which is roughly 1.1 and 0.6 $\pm$ 0.2 eV for LaOBiS$_2$ and LaOBiSSe respectively.   The reduction of the band gap due to Se-substitution is also verified by ($\alpha${\it h}$\nu$/s)$^2$ measurements \cite{MiuraOptical} on LaOBiS$_2$ and LaOBiSSe shown in Figs. 2(d) and (e), where direct-transition-type optical absorption was observed for 1.0 $\pm$ 0.1 eV for LaOBiS$_2$ and 0.5 $\pm$ 0.1 eV for LaOBiSSe. 
%Thus the photoemission results seems to be referencing the bulk nature of material.

\begin{figure}
\includegraphics{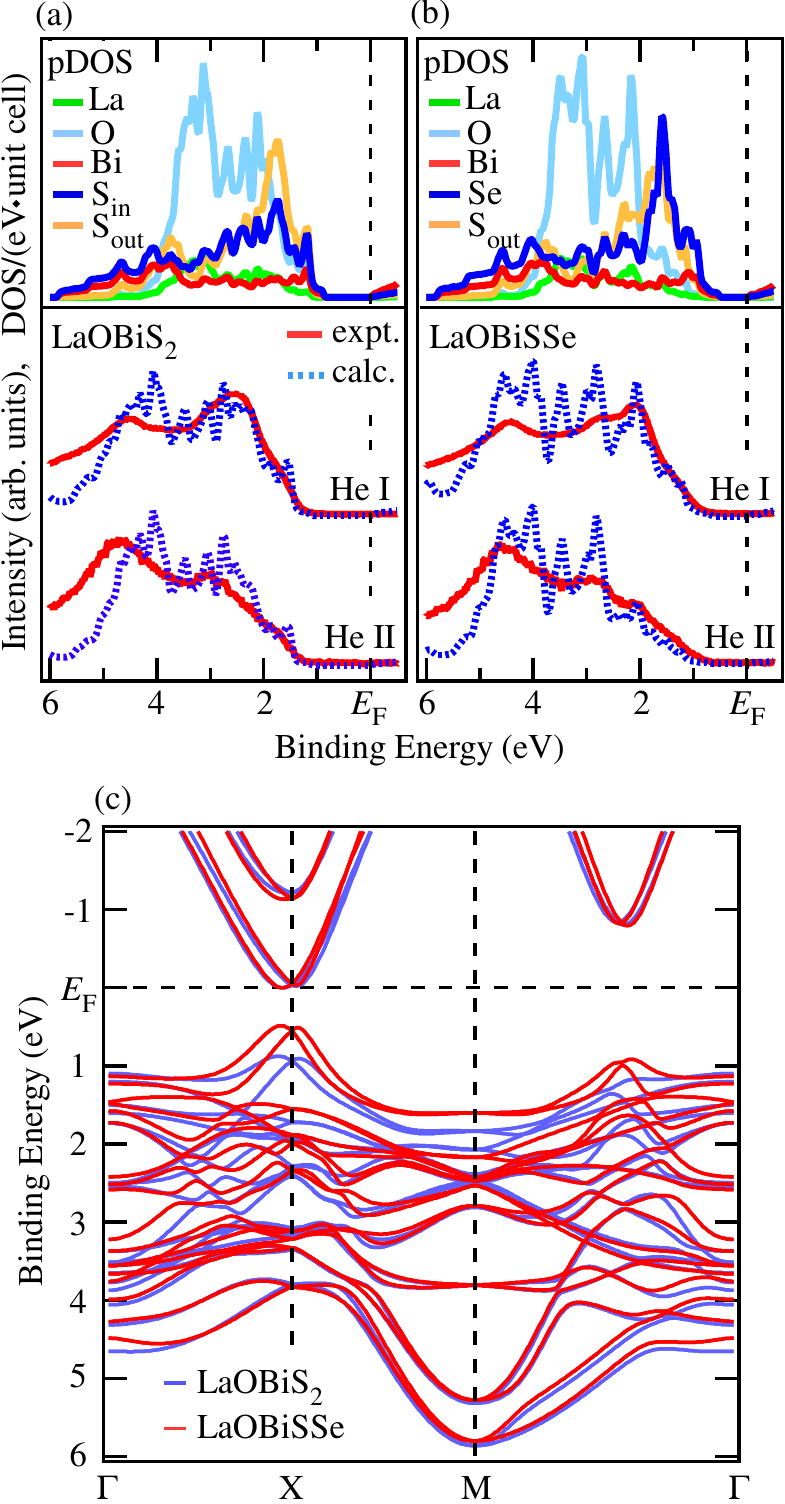}%
\caption{(a) and (b) Top: Calculated pDOS for LaOBiS$_2$ and LaOBiSSe. Bottom: Photoemission data (red lines) and simulated spectra (blue dashed lines) based on calculated pDOS and photoionization cross section.  (c) Calculated band structure along high symmetry lines for LaOBiS$_2$ (blue) and LaOBiSSe (red).}
\end{figure}

	In order to discuss the origin of the observed change in the electronic structure, we have performed band structure calculations including spin-orbit interaction.  Upper panels of  Figs. 3(a) and (b) show calculated partial density of states (pDOSs) for LaOBiS$_2$ (a) and LaOBiSSe (b), where S$_{in}$ denotes in-plane S atom and S$_{out}$ is out-of-plane S atom. Following earlier experimental and theoretical works \cite{JAP2, GWang}, we have assumed that Se atoms occupy only in-plane S sites.  Bottom of Figs. 3(a) and (b) are experimental spectra (red lines), overlapped with simulated photoemission spectra (blue dashed lines) where each pDOS was multiplied by photoionization cross section \cite{Goldberg, Lindau} and extended 1.3 times in energy for comparison.   For calculated results, the position of $E_F$ was set at the bottom of the conduction band, in order to see the correspondence with the photoemission results in Fig. 2.  The simulated spectra successfully captures the characteristics of Chalcogen-substitution induced change in the valence band. Namely, (i) the peak structure $\sim$2.5 eV in LaOBiS$_2$ splits into $\sim$2.5 eV and $\sim$2 eV in LaOBiSSe (ii) the top of the valence band becomes closer to $E_F$ and the band gap decreases by Se-substituion.  Thus it can be said that the first principles calculations would give reliable prediction in these materials.  
	
	According to the calculations shown in Figs. 3(a) and (b), the changes in the experimental spectra are mainly attributed to the pDOS of in-plane chalcogen atom.  As shown in Fig. 3(c), it is expected that the top of the valence band is largely affected by the substitution of S with Se while the bottom of conduction band is less affected although the dispersion becomes slightly steeper in LaOBiSSe (Estimated effective masses ({\it m$^*$}) for the bottom of the conduction bands up to 0.1 eV are 0.16 and 0.11 for LaOBiS$_2$ and LaOBiSSe, respectively).  Such a change in the band structure would contribute thermoelectric property especially at high temperature, whereas it would not fully explain the reduction of the resistivity in Fig. 1 since the chemical potential seems to be pinned at the bottom of the conduction band and the population of Fermi-Dirac function at binding energy of 0.5-1.0 eV hardly changes below room temperature (see inset of Fig. 2(c)).  Therefore, chalcogen substitution certainly modifies the band structure but the change would not fully account for the improvement of thermoelectric property.  Next we focus on the Se-substitution effect on the local structure.

Figure 4 shows Fourier transforms (FT) of  $k$$^2$-weighted EXAFS oscillations extracted from the Bi L$_{3}$-edge x-ray absorption spectra measured on LaOBiS$_2$ and LaOBiSSe samples at $T$ = 30 K. The inset includes EXAFS oscillations ($k^2$-weighted) as a function of photoelectron wave vector $k$. The FTs are performed using a gaussian window with the $k$ range being 3.5-15 \AA$^{-1}$ and corrected for the phase shifts. The EXAFS oscillations and the FT peak structures are largely different revealing distinct local structure of the two compounds. 

%Figure 5
\begin{figure}
\includegraphics[width=8cm]{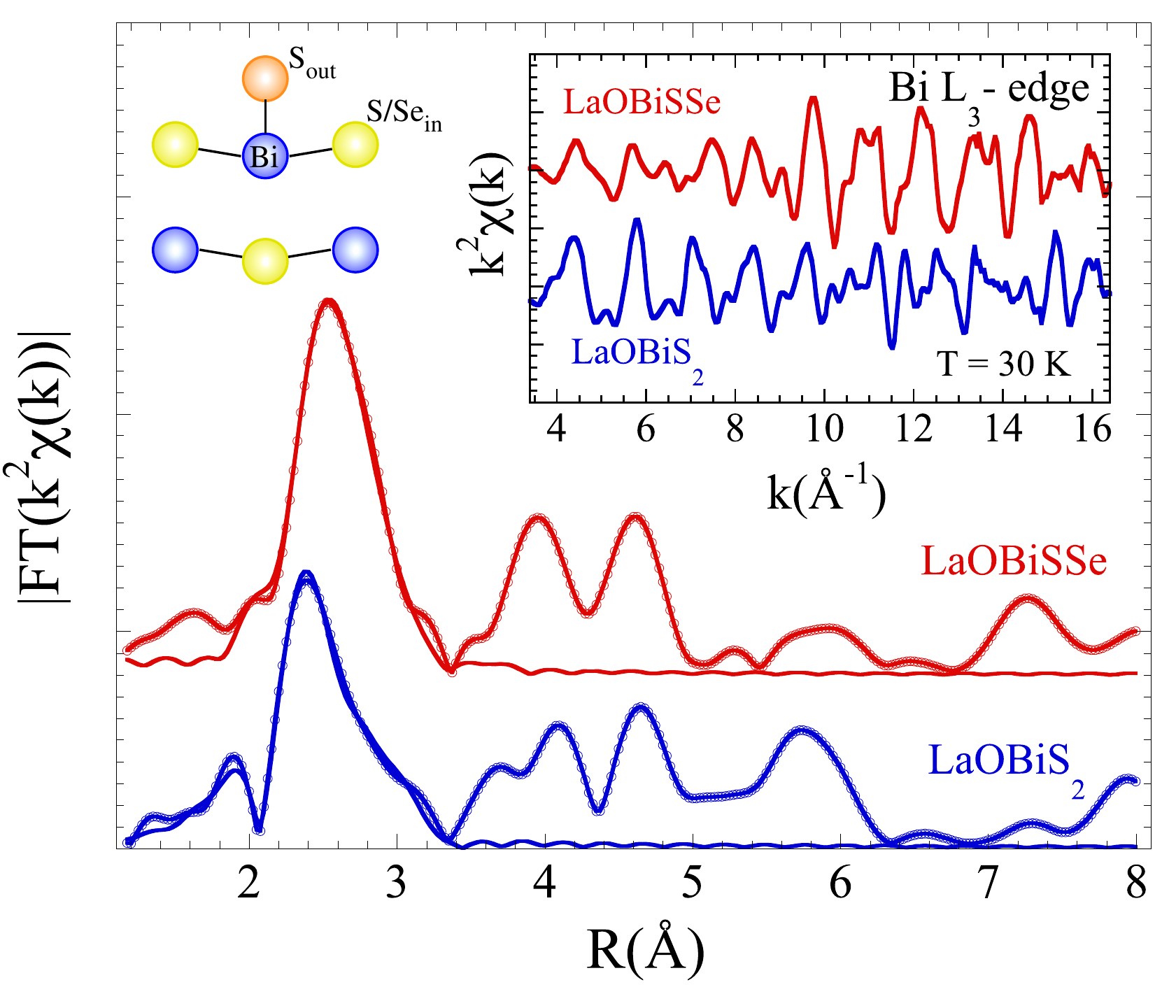}
\caption{Fourier transform (FT) magnitudes of the Bi $L_3$-edge EXAFS of LaOBiS$_2$ and LaOBiSSe. Model fits to the FTs are also shown as solid lines. EXAFS oscillations of the two samples and schematic view of atoms surrounding Bi are displayed as inset.}
\label{}
\end{figure}

In LaOBiS$_2$ structure, the Bi atom is coordinated with one sulphur atom in axial direction (S$_{out}$), four in-plane sulphur atoms (S$_{in}$), and the S$_{in}$ atom of the adjacent BiS$_{2}$ layer. Information on all these Bi-S distances is contained in the main peak structure between 1.5 and 3.5 \AA$ $ in the FTs of Bi L$_{3}$-edge EXAFS. The contributions of distant atoms are mixed with multiple scatterings and visible beyond the first peak structure. To determine the local structure parameters, the EXAFS oscillations were modeled by the general equation \cite{Bunker} based on single-scattering approximation.  Following earlier EXAFS study and theoretical calculations \cite{paris, paris2, poly1}, we used two different in-plane Bi-S$_{in}$ distances for LaOBiS$_2$.  The model contains four shells including one out-of-plane Ch$_{out}$ atom (Bi-Ch$_{out}$ distance), four in-plane Ch$_{in}$ atoms at two different distances (two Bi-Ch$_{in}$ distances) and one Bi-Ch$_{in}$ distance between the two BiCh$_2$-layers. For LaOBiSSe, considering the fact that Se preferentially occupies the S$_{in}$ site \cite{JAP2}, Se occupation at S$_{out}$ site was neglected. We have fixed the passive electrons reduction factor, $S_0^2$ = 0.95 for the analysis of the Bi L$_3$-edge EXAFS, while the photoelectron energy zero, $E_0$ was set to zero after analysing different scans. The only parameters refined here are the local bondlengths and the associated $\sigma^2_i$, measuring the mean square relative displacement (MSRD) of the considered bond distances. The FEFF8 code \cite{Feff} was used for the calculation of scattering amplitudes and the WINXAS package \cite{winxas} was used for the EXAFS model fits. The fit $k$-range was 3.5-15 \AA$^{-1}$ while $R$-range was 1.5-3.5 \AA$ $. Therefore, the number of independent points 2$\Delta k\Delta R/\pi$ was $\sim$14 for the model fits in which the fit parameters were 8 to 10.  The model fits are included in Fig. 4 as solid lines.
%
%\begin{equation}
%\chi(k)= \sum_{i}\frac{N_{i}S_{0}^{2}}{kR_{i}^{2}}f_{i}(k,R_{i})
%e^{-\frac{2R_{i}}{\lambda}} e^{-2k^{2}\sigma_{i}^{2}}
%\sin[2kR_{i}+\delta_{i}(k)],\nonumber
%\end{equation}

%\noindent where $N_{i}$ is the number of neighbouring atoms at distances $R_{i}$ from the photoabsorbing atom, $\delta_{i}$ is the phase shift, $f_{i}(k,R_{i})$ is the backscattering amplitude, $\lambda$ is the photoelectron mean free path, and $\sigma_{i}^{2}$ is the correlated Debye-Waller factor (DWF) measuring the mean square relative displacement (MSRD) of the pair of atoms (photoabsorber-backscatter pairs). $S_{0}^{2}$ is the so-called passive electrons reduction factor  \cite{}. As a starting model we have used the structure deduced from x-ray diffraction measurements on LaOBiS$_2$  (for the Bi-S1 and Bi-S2 bonds) and LaOBiSSe. (for Bi-Se bonds contributions) \cite{miz_SciRep}.  

%Table 1
\begin{table}
\caption{\label{tab:table1}Bond distances (R) and mean square relative displacements ($\sigma^{2}$) of LaOBiS$_2$ and LaOBiSSe determined by Bi L$_3$-edge EXAFS ($T$ = 30 K).  Maximum uncertainty in distance determination is about $\sim$0.01 \AA, and in the corresponding $\sigma^{2}$ is about
$\sim$0.001 \AA$^{2}$. Bi-Ch$_{in}^{1}$ and Bi-Ch$_{in}^{2}$ are two in-plane Bi-Ch$_{in}$ distances while Bi-Ch$_{in}^{i}$ is the distance between the BiCh$_2$-layers.}
%\begin{hline}
%\begin{ruledtabular}
\begin{tabular}{ccccccccccc}\hline
\hline
&&LaOBiS$_2$&&&LaOBiSSe&&\\
\hline
&R(\AA)&$\sigma^{2}$(\AA$^{2}$)&&&R(\AA)&$\sigma^{2}$(\AA$^{2}$)\\
Bi-S$_{out}$$^{ }$&2.47&0.003&&&2.48&0.004&\\
Bi-S/Se$_{in}^{1}$&2.68&0.003&&&2.82&0.003&\\
Bi-S/Se$_{in}^{2}$&3.09&0.005&&&3.04&0.006&\\
Bi-S/Se$_{in}^{i}$&3.42&0.006&&&3.47&0.009&\\
\hline
\hline
\end{tabular}
%\end{ruledtabular}
\end{table}

The local structure parameters obtained for two compounds are summarized in table 1. The results on LaOBiS$_2$ are consistent with earlier EXAFS work on the local structure of F-doped system\cite{paris} and PDF analysis \cite{Athauda} showing Bi-S$_{in}$ bondlength splitting. This confirms that BiS$_2$ layer is largely distorted in LaOBiS$_2$. On the other hand, LaOBiSSe contains two Bi-Se$_{in}$ distances separated by $\sim$0.2 \AA.  From table 1, it is apparent that local distortion in LaOBiSSe is substantially lower with respect to that in LaOBiS$_2$.  Therefore, it is lilkely that structural instability in LaOBiS$_2$, characterized by polytypism, is partially suppressed in LaOBiSSe. Apart from other factors like microstructure, we think this could be the main origin of the change in the resistivity and hence the thermoelectric properties through chalcogen substitution, especially at lower temperature range.

The present EXAFS and earlier experimental works \cite{Athauda, Sagayama, paris} indicate that Bi atoms in LaOBiS$_2$ are in off-center position in S$_{in}$ square lattice.  Such a lowering of the structural symmetry has been intensively studied in post-transition metal oxides and chalcogenides \cite{Walsh}, where it turned out that the lone pair becomes active when the energy level of chalcogen $p$-orbitals are close enough to that of $s$-orbital in post-transition metal.  If the hybridization between those two are strong enough, the post-transition metal atoms choose to take off-center position, forming $s$-$p$ hybridization to obtain energy gain in the electronic system.  We think this idea can be also applicable in the current BiCh$_2$ system, as is the case of Bi$_2$O$_3$, Bi$_2$S$_3$, and Bi$_2$Se$_3$ \cite{Walsh2}.  Indeed, the overall energy position of calculated Se 4$p$ pDOS is higher compared with that of S 3$p$ pDOS in Figs. 3(a) and (b).   Therefore, although detailed bond analysis\cite{Walsh2} on this series of material should be important to conclude, the manipulation of the lone pair activity by chalcogen subsituion could reduce the local distortion in the conducting layer and recover the conductivity.  In the parent compound LaOBiS$_2$, the contribution of electrons to thermal conductivity ($\kappa_e$) is small compared with that of phonons ($\kappa_{ph}$) \cite{JAP1}.  The drastic decrease of resistivity increases $\kappa_e$ but it seems to be cancelled by the decrease of $\kappa_{ph}$ with help of possible rattling motion of Bi \cite{Lee}.  As a result, the $ZT$ value would have benefitted from chalcogen substitution.   Such a control of the local structural distortion in the compounds with the lone pair (such as Bi, Pb, Sn) can be benefical for further improvement of thermoelectric property.

\section{CONCLUSION}
	In summary, we have studied Se-substitution effect on the electronic- and the local structure of LaOBiCh$_2$ system.  It turned out that the top of the valence band is significantly influenced by the substitution where pDOS of in-plane chalcogen atom is dominant.  We have also observed that the local structural distortion in the Bi-chalcogen plane is suppressed by Se-substitution, which is in accordance with the lone pair activity of the system.  It is suggested that controlling the local structural distortion can help enhancing the functionality of thermoelectric materials.

\begin{acknowledgments}
	We thank ESRF staff for support in the EXAFS data collection.  The authors would like to thank S. Onari for the use of workstation.  K. T. and T. W. would like to acknowledge the hospitality at the Sapienza University of Rome.  This research was partially supported by the Program for Promoting the Enhancement of Research University from MEXT, the Program for Advancing Strategic International Networks to Accelerate the Circulation of Talented Researchers from JSPS (R2705), and JSPS KAKENHI (Nos. 15H03691, 16H04493).  This work is a part of the executive protocol of the general agreement for cooperation between the Sapienza Unversity of Rome and Okayama University, Japan.

\end{acknowledgments}


\begin{thebibliography}{39}
\bibitem{mono}
	The electronic structure under the monoclinic crystral structure has been calculated by ref. 22, where splitting of bands in the order of 0.1 eV has been predicted.
\end{thebibliography}
\end{document}